\documentclass[aps,prd,floatfix,reprint,amsmath,amssymb,superscriptaddress]{revtex4-2}

\usepackage{amsfonts}
\usepackage{mathrsfs}
\usepackage{amsmath}% needed for subequations
\usepackage{color}
\usepackage{natbib}
\usepackage{graphicx}
\usepackage{bm}% bold maths
\usepackage{amssymb}
\usepackage{xspace}
\usepackage{epstopdf}
\usepackage{dcolumn}% Align table columns on decimal point
\usepackage{longtable}
\usepackage{multirow}
\usepackage{textcomp}

\usepackage{titlesec}
\titleformat{\section}[block]{\bfseries\filcenter}{}{1em}{}

\usepackage[colorlinks=true, letterpaper=true, pdfstartview=FitV, linkcolor=blue, citecolor=blue, urlcolor=blue]{hyperref}

\makeatother
\begin{document}
%\title{Unveiling the Anharmonic-induced Phonon Softening and Thermal-Electrical Transport Properties in Nodal-Line Semimetal ZrSiS}
%\title{Anomalous thermal transport and exceptional electrical conductivity in nodal-line semimetal ZrSiS}
\title{Anharmonicity-driven avoided phonon crossing and anomalous thermal transport in the nodal-line semimetal ZrSiS}

\author{Xin Jin}
\affiliation{College of Physics and Electronic Engineering, Chongqing Normal University, Chongqing 401331, China}

\author{Qingqing Zhang}
\affiliation{College of Physics and Electronic Engineering, Chongqing Normal University, Chongqing 401331, China}

\author{Dengfeng Li}
\affiliation{School of Science, Chongqing University of Posts and Telecommunications, Chongqing, 400065, China}

\author{Zhenxiang Cheng}
\affiliation{Institute for Superconducting and Electronic Materials Faculty of Engineering and Information Sciences University of Wollongong Innovation Campus, Squires Way, North Wollongong, NSW 2500, Australia}

\author{Jianli Wang}
\affiliation{College of Physics and Electronic Engineering, Chongqing Normal University, Chongqing 401331, China}

\author{Xuewei Lv}
\affiliation{College of Materials Science and Engineering, Chongqing University, Chongqing 400044, P. R. China.}

\author{Xiaoyuan Zhou}
\affiliation{College of Physics, Chongqing Key Laboratory for Strongly Coupled Physics, Chongqing University, Chongqing 401331, P.R. China}
\affiliation{College of Physics, and Center of Quantum Materials and Devices, Chongqing University, Chongqing 401331, China}

\author{Rui Wang}
\affiliation{College of Physics, Chongqing Key Laboratory for Strongly Coupled Physics, Chongqing University, Chongqing 401331, P.R. China}
\affiliation{College of Physics, and Center of Quantum Materials and Devices, Chongqing University, Chongqing 401331, China}

\author{Xianyong Ding}
\email{dxy$_$vasp@163.com}
\affiliation{College of Physics and Electronic Engineering, Chongqing Normal University, Chongqing 401331, China}

\author{Peng Yu}
\email{pengyu@cqnu.edu.cn}
\affiliation{College of Physics and Electronic Engineering, Chongqing Normal University, Chongqing 401331, China}

\author{Xiaolong Yang}
\email{yangxl@cqu.edu.cn}
\affiliation{College of Physics, Chongqing Key Laboratory for Strongly Coupled Physics, Chongqing University, Chongqing 401331, P.R. China}
\affiliation{College of Physics, and Center of Quantum Materials and Devices, Chongqing University, Chongqing 401331, China}

\begin{abstract}
Understanding thermal and electrical transport in topological materials is essential for advancing their applications in quantum technologies and energy conversion. Herein, we employ first-principles calculations to systematically investigate phonon and charge transport in the prototypical nodal-line semimetal ZrSiS. The results unveil that anharmonic phonon renormalization results in the pronounced softening of heat-carrying phonons and suppressed lattice thermal conductivity ($\kappa_{\rm L}$). Crucially, anharmonic effects are found to noticeably weaken Zr-S interactions, triggering avoided-crossing behavior of low-frequency optical phonons. The combination of phonon softening and avoided crossing synergistically reduces phonon group velocities, yielding a 16\% suppression in $\kappa_{\rm L}$ along the $c$-axis at room temperature. Contrary to conventional metals, we discover that the lattice contribution to thermal conductivity in ZrSiS is abnormally large, even dominating heat conduction along the $c$-axis. This unusual behavior results in a substantial deviation of the Lorenz number from the Sommerfeld value---exceeding it by up to threefold---thereby challenging the validation of standard Wiedemann-Franz law for thermal conductivity estimation. Moreover, our calculations demonstrate that ZrSiS exhibits exceptional electrical conductivity, attributed to its topological electronic Dirac states that account for both high Fermi velocities and weak electron-phonon coupling. This study provides critical insights into the electrical and thermal transport mechanisms in ZrSiS and highlights the importance of anharmonic effects in the lattice dynamics and thermal transport of metallic materials.

\end{abstract}

% \pacs{73.20.At, 71.55.Ak, 74.43.-f}

%Use showkeys class option if keyword %display desired

\maketitle

\section{Introduction}
Topological semimetals have garnered significant attention over the past decades because of their fundamental and practical importance. Among the diverse family of topological semimetals, nodal-line semimetals (NLSMs) have emerged as a vibrant research frontier in condensed matter physics, exhibiting unique electronic structures and novel quantum transport phenomena \cite{burkov2011topological, fang2016topological, yu2017topological, yang2018symmetry, yang2022quantum, rui2018topological}. The uniqueness of these materials lies in the one-dimensional topological nodal lines or rings within their electronic band structures, which lay the foundation for their novel electronic transport properties, including the dissipationless electronic transport \cite{rui2018topological, hu2019transport, syzranov2017electron}, electronic correlations \cite{shao2020electronic, liu2017correlation}, optical conductivity \cite{pronin2021nodal}, and quantum anomaly \cite{burkov2018quantum}. Recent advances have enabled the successful prediction and experimental realization of numerous NLSMs \cite{chang2019realization, fang2015topological, he2018type, xu2017topological, hosen2017tunability, bian2016topological, qiu2019observation, wang2021spectroscopic, lv2021experimental,song2020photoemission}. Among these, ZrSiS stands out as a prototypical material, distinguished by a unique coexistence of symmetry-protected nodal lines and nodal surfaces \cite{schoop2016dirac, fu2019dirac, neupane2016observation}. This material offers distinct experimental advantages, as it can be readily synthesized from relatively abundant and non-toxic elements through chemical vapor transport \cite{schoop2016dirac}. More notably, ZrSiS exhibits highly anisotropic, large non-saturating magnetoresistance (MR) even under strong magnetic fields \cite{novak2019highly, lv2016extremely, singha2017large, ali2016butterfly}, making it an exceptional platform for exploring novel quantum phenomena and potential device applications. 

Given that thermal conductivity ($\kappa$) is critical to fundamental science and many applications, it is imperative to develop a comprehensive understanding of the thermal transport properties of ZrSiS. For metals or semimetals, thermal transport was generally believed to be predominantly governed by free electrons, and their contribution to thermal conductivity ($\kappa_{\rm e}$) is conventionally estimated through the Wiedemann-Franz Law (WFL) \cite{chester1961law}. According to this law, $\kappa_{\rm e}$ can be expressed as $\kappa_{\rm e}$ = $L \sigma T$, where $L$ represents the Lorenz number, $\sigma$ denotes the experimentally measured electrical conductivity, and $T$ stands for the temperature. This relationship provides a crucial link between the thermal and electrical properties of metals, enabling the determination of thermal conductivity through electrical conductivity measurements. However, it has been reported that in many metallic systems, such as tungsten \cite{chen2019understanding}, beryllium \cite{chen2024origin}, and $\theta$-TaN \cite{kundu2021ultrahigh}, the estimated thermal conductivity from WFL using the Sommerfeld value of $L_{0}$ = 2.44 $\times$ 10$^{-8}$W$\Omega$K$^{-2}$ \cite{makinson1938thermal,klemens1986thermal}, shows substantial deviations from experimental measurements. This discrepancy has been attributed to the neglect of lattice component ($\kappa_{\rm L}$) of thermal conductivity, which was identified to be anomalously large and even dominant in thermal transport for these systems \cite{chen2019understanding,chen2024origin,kundu2021ultrahigh}. In this scenario, quantifying the contributions of electrons and phonons' contributions to $\kappa$ in metallic systems, along with accurate knowledge of Lorenz number, is of paramount significance. Solving this puzzle not only facilitates the deeper understanding of microscopic mechanisms governing heat transfer in metals but also provides critical guidance for experimental measurements. More importantly, for topological semimetals like ZrSiS, the fundamental connection between their non-trivial topological band structures and thermal transport properties remains to be elucidated.

Recent advancements in first-principles calculation based on density functional theory (DFT), integrated with the Boltzmann transport equation (BTE), have enabled precise prediction of both phononic and electronic contributions to $\kappa$ \cite{lindsay2018survey,yang2021indirect,yang2021tuning}. In metals, the intrinsic $\kappa_{\rm L}$ is governed by anharmonic phonon-phonon interactions and the scattering of phonons by electron (ph-el), while the intrinsic $\kappa_{\rm e}$ is limited by the scattering of electrons by phonons (el-ph). Within the framework of harmonic approximation (HA), the phonon linewidths and $\kappa_{\rm L}$ have been accurately predicted and reasonably understood for numerous materials \cite{chen2019understanding,chen2024origin,kundu2021ultrahigh}. However, emerging studies \cite{xia2020particlelike,yang2022reduced,han2022raman,ding2024anharmonicity,yue2025interlayer} have revealed that temperature-induced anharmonic phonon renormalization (APR) effect plays a crucial role in determining the phonon energy, lifetimes, and $\kappa_{\rm L}$, particularly in strongly anharmonic materials. As metallic systems generally have strong lattice anharmonicity, it is thus important to understand the physics of heat transport in ZrSiS by incorporating lattice dynamical effects beyond the HA.

In this work, we conduct a comprehensive investigation of the intrinsic phonon and electron transport properties of ZrSiS using first-principles calculations coupled with the BTEs of phonon and electron. Our results show that the APR induces phonon softening and avoided-crossing behavior, significantly decreasing phonon group velocities and consequently suppressing the $\kappa_{\rm L}$ along the $c$-axis by 16\% at room temperature (RT). By incorporating both APR effects and ph-el scattering, we predict the anisotropic $\kappa_{\rm L}$ of ZrSiS and demonstrate that $\kappa_{\rm L}$ is anomalously large compared to its electronic contribution, even dominating heat conduction along the $c$-axis. This leads to a substantial deviation of the Lorenz number from the Sommerfeld constant, exceeding it by up to a factor of three. Furthermore, we compute the phonon-limited electrical conductivity, which aligns well with experimental measurements \cite{hussain2020electron,singha2017large}, and uncover exceptional electrical conductivity in ZrSiS. This outstanding performance arises from high Fermi velocities and weak el-ph coupling, both attributed to its topological Dirac electronic states. Our work not only elucidates the fundamental thermal transport mechanisms in ZrSiS but also underscores the critical role of APR effect in understanding heat conduction in metallic systems.

\section{Methodology}
By solving the phonon BTE, the lattice thermal conductivity tensor ($\kappa_{\rm L}^{\alpha\beta}$) can be obtained as \cite{lindsay2010flexural}
\begin{equation}\label{kappal}
    \kappa_{\rm L}^{\alpha\beta} = \sum_{p \mathbf{q}}C_{p \mathbf{q}}v_{p \mathbf{q}}^{\alpha} \otimes v_{p \mathbf{q}}^{\beta}\tau_{p \mathbf{q}}
\end{equation}
where $\alpha$ and $\beta$ are Cartesian axes, and $C_{p \mathbf{q}}$, $\nu_{p \mathbf{q}}$, and $\tau_{p \mathbf{q}}$ represents the heat capacity, phonon group velocity, and phonon lifetime, respectively, of phonons from branch $p$ with wave vector $\mathbf{q}$. Here, $\tau_{p \mathbf{q}}$ is obtained from an iterative solution of the linearized BTE of phonons starting with the relaxation time approximation (RTA) \cite{omini1997heat}. Within the RTA, the phonon scattering rate, inverse of $\tau_{p \mathbf{q}}$, is calculated as a Matthiessen sum of contributions from anharmonic three-phonon (3ph, $1/\tau^{\rm 3ph}_{p \mathbf{q}}$), phonon-isotope ($1/\tau^{\rm iso}_{p \mathbf{q}}$), and ph-el ($1/\tau^{\rm ph-el}_{p \mathbf{q}}$) scattering processes. Their full expressions can be found elsewhere \cite{li2014shengbte,han2022fourphonon,feng2017four,chen2019understanding,yang2022reduced,ding2024anharmonicity,wei2024tensile}. 

The electrical conductivity tensor ($\sigma^{\alpha\beta}$) and electronic thermal conductivity tensor ($\kappa_{\rm e}^{\alpha\beta}$) can be calculated within the framework of the linearized electron BTE using the following expressions \cite{li2015electrical,madsen2006program}
\begin{equation}\label{eleSigma}
\sigma^{\alpha\beta} = \frac{2e^{2}}{N_{\mathbf{k}}V_0 k_{B} T} \sum_{n\mathbf{k}} f_{n\mathbf{k}}^{0}(1-f_{n\mathbf{k}}^{0})\mathbf{v}_{n\mathbf{k}}^{\alpha} \otimes \mathbf{F}_{n\mathbf{k}}^{\beta},
\end{equation}

\begin{equation}\label{kappae}
\begin{split}
\kappa_{\rm e}^{\alpha\beta} =& \frac{2}{N_{\mathbf{k}} V_0 k_{B} T^{2}} \sum_{n\mathbf{k}} f_{n\mathbf{k}}^{0}(1-f_{n\mathbf{k}}^{0}) \\
&\times (E_{n\mathbf{k}} - E_{f})^{2} \mathbf{v}_{n\mathbf{k}}^{\alpha} \otimes \mathbf{F}_{n\mathbf{k}}^{\beta} - T (\sigma^{\alpha\beta}S^{\alpha\beta})^2/\sigma^{\alpha\beta},
\end{split}
\end{equation}
and 
\begin{equation}\label{sigmas}
\begin{split}
\sigma^{\alpha\beta} S^{\alpha\beta} =& \frac{2}{N_{\mathbf{k}} V_0 k_{\rm B} T^{2}} \sum_{n\mathbf{k}} f_{n\mathbf{k}}^{0}(1-f_{n\mathbf{k}}^{0}) \\
& \times (E_{n\mathbf{k}} - E_{f}) \mathbf{v}_{n\mathbf{k}}^{\alpha} \otimes \mathbf{F}_{n\mathbf{k}}^{\beta},
\end{split}
\end{equation}
where $e$ is the elementary charge, $k_{\rm B}$ is the Boltzmann constant, $n$ and $\mathbf{k}$ is the band index and wave vector, $N_{\mathbf{k}}$ is the number of uniformly sampled $\mathbf{k}$ points, $f_{n\mathbf{k}}^{0}$ is the equilibrium electron Fermi-Dirac distribution. $E_{n \mathbf{k}}$ and $E_{f}$ are the electronic energy and Fermi energy. $\mathbf{v}_{n\mathbf{k}}$ is the electron group velocity. $\mathbf{F}_{n\mathbf{k}}$ is the electron mean free path limited by el-ph scattering, i.e., $\mathbf{F}_{n\mathbf{k}}$=$v_{n\mathbf{k}}\tau_{n\mathbf{k}}^{\rm el-ph}$. $\tau_{n\mathbf{k}}^{\rm el-ph}$ is the phonon-limited electron lifetime, which is accurately solved with an iterative scheme. 

Solving the numerical phonon BTE requires harmonic and anharmonic interatomic force constants (IFCs), which are obtained from DFT calculations using the Vienna $Ab$ $initio$ Simulation Package (VASP) \cite{kresse1993ab, kresse1996efficient}. The exchange-correlation functional is treated within the generalized gradient approximation as parameterized by the Perdew-Burke-Ernzerhof functional \cite{perdew1996generalized}, with a plane-wave cutoff energy of 500 eV. The ground-state 2nd, 3rd, and 4th-order IFCs calculations are performed using the finite-difference approach with a 3$\times$3$\times$3 supercell and a 3$\times$3$\times$2 Monkhorst-Pack $\mathbf{k}$ grid, as implemented in Phonopy \cite{togo2023first}, Thirdorder \cite{li2014shengbte}, and Fourthorder \cite{han2022fourphonon}, respectively. Cutoff radii of interatomic interactions are set to 0.5 and 0.3 {\AA} for the 3rd and 4th-order IFCs, respectively. To account for APR effect, temperature-dependent (TD) IFCs are extracted using the temperature-dependent effective potential (TDEP) method \cite{bottin2020tdep}. To generate atomic trajectories for inputs of the TDEP, ab initio molecular dynamics simulations are carried out in a 4$\times$4$\times$4 supercell with 384 atoms at finite temperatures for 15000 steps with the canonical (NVT) ensemble and a time step of 1 fs. The convergence tests of the cutoff radius for the 2nd and 3rd-order TD IFCs are provided in Fig. S1(b) of the Supplemental Material (SM) \cite{SM}. The extracted TD IFCs at several different temperatures display only a weak dependence on $T$, as evidenced by the negligible changes in the resulting phonon spectra across the studied range in Fig. S2 of the SM \cite{SM}. Thus, to achieve a balance between computational accuracy and efficiency, we utiliz the IFCs at 300 K to calculate the final $\kappa_{\rm L}$ over the entire $T$ range. 

The electron energies are calculated within density functional perturbation theory (DFPT) by the Quantum Espresso package \cite{giannozzi2009quantum}, and the electron-ion interactions are described by ultrasoft pseudopotentials \cite{vanderbilt1990soft,dal2014pseudopotentials} with a cutoff energy of 50 Ry for wave functions. To calculate the ph-el scattering rates, the ph-el coupling matrix elements are initially obtained on coarse electron grids with 12$\times$12$\times$8 $\mathbf{k}$-point and 3$\times$3$\times$2 $\mathbf{q}$-point grid and then interpolated onto a dense 48$\times$48$\times$32 for both $\mathbf{k}$ and $\mathbf{q}$ grid using the maximally localized Wannier function basis as implemented in the EPW package \cite{ponce2016epw}. With the same $\mathbf{q}$ grid, $\kappa_{\rm L}$ is determined by solving the phonon BTE with an iterative scheme using the modified ShengBTE package \cite{li2014shengbte}. For the calculation of phonon-limited electrical transport properties, final grids of 85$\times$85$\times$85 for both $\mathbf{k}$ and $\mathbf{q}$ points are adopted to calculate the el-ph scattering rates and solve the linearized electron BTE via the Perturbo code \cite{zhou2021perturbo}. As demonstrated in the SM, this computational setup guarantees the convergence of $\kappa_{\rm L}$, $\kappa_{\rm e}$, and $\sigma$. The interatomic bonding strengths are characterized by calculating the integrated crystal orbital Hamilton population (ICOHP) \cite{maintz2013analytic,dronskowski1993crystal,deringer2011crystal} using the LOBSTER code \cite{maintz2016lobster}.

\section{Results and Discussion} 
\subsection{Anisotropic crystal structure and bonding heterogeneity} 
\begin{figure}[!t]
	\includegraphics[width=\linewidth]{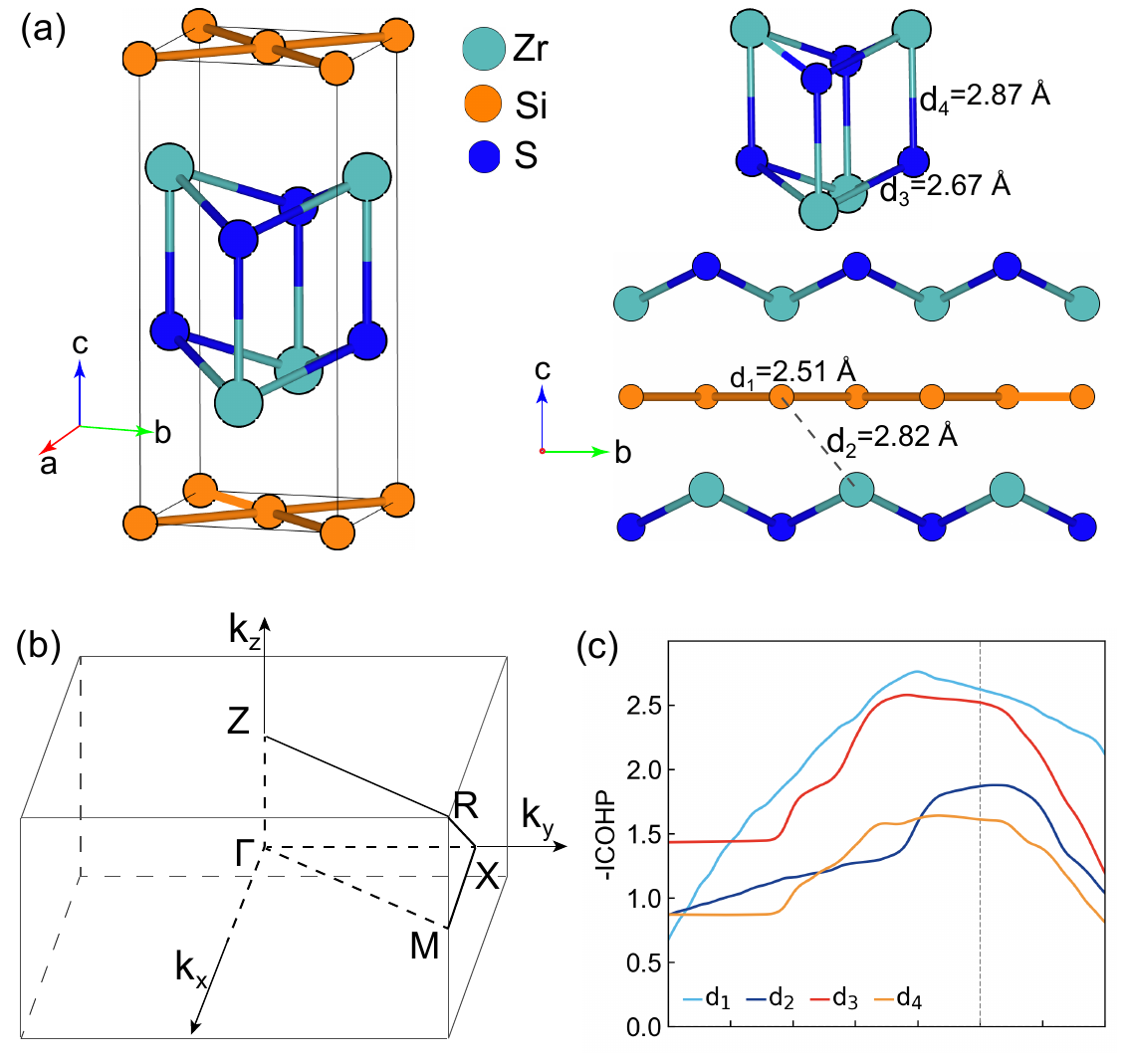}
	\caption{(a) Lattice structure of ZrSiS, the bond length for Zr-Si, Zr-S, and Si-Si are marked. (b) The first Brillouin zone of ZrSiS. (c) ICOHP for various chemical bonds represented in (a).}
\label{fig1}
\end{figure}

ZrSiS crystallizes in the Matlockite tetragonal structure with a space group of $P4/nmm$ (No. 129), as illustrated in Fig.~\ref{fig1}(a). In this structure, Zr and S atoms occupy the Wyckoff positions of 2$c$ (0, 0.5, $u$) with point group symmetry of $4mm$, with $u$ = 0.37853033 and 0.730605155, respectively, while Si atoms occupy the Wyckoff position of 2$a$ (0, 0, 0) with point group symmetry of $\overline{4}2m$. The optimized lattice parameters are $a$ = $b$ = 3.55 $\rm{\AA}$ and $c$ = 8.14 $\rm{\AA}$, showing good agreement with experimental results \cite{onken1964silicid}. ZrSiS exhibits a distinctive layered structure characterized by a square network of Si atoms within the $ab$ plane, intercalated with alternating layers of Zr and S atoms. The different layers are stacked together through covalent bonds rather than van der Waals interactions \cite{zhou2017lattice}, making it impossible to obtain nanometer-thick flakes from the bulk material via mechanical exfoliation. In the $ab$ plane, each Si atom is covalently bonded to four equivalent Si atoms with a bond length of 2.51 $\rm{\AA}$ ($d_{1}$), while the Zr and S atoms are covalently connected with a bond length of 2.67 $\rm{\AA}$ ($d_{3}$) and form zigzag chains. Along the $c$-axis, each Si atom is connected to four equivalent Zr atoms with a bond length of 2.82 $\rm{\AA}$ ($d_{2}$), while Zr and S atoms are bonded at a distance of 2.87 $\rm{\AA}$ ($d_{4}$). The significantly longer bond distance between interlayer atomic pairs ($d_{2}$ and $d_{4}$) compared to those within the intralayers ($d_{1}$ and $d_{3}$) indicate the much weaker interlayer bonding strength between different atom pairs. This is further supported by the ICOHP values shown in Fig.~\ref{fig1}(c), revealing that the ICOHP values of $d_{1}$ and $d_{3}$ around the Fermi energy are markedly larger than those of $d_{2}$ and $d_{4}$. These results indicate that ZrSiS exhibits strong atomic bonding heterogeneity along the $c$-axis and within the $ab$ plane, which is expected to result in highly anisotropic transport properties as elaborated later.

\subsection{Phonon softening and avoided crossing driven by strong anharmonicity} 

\begin{figure*}[!tbp]
	\includegraphics[width=\linewidth]{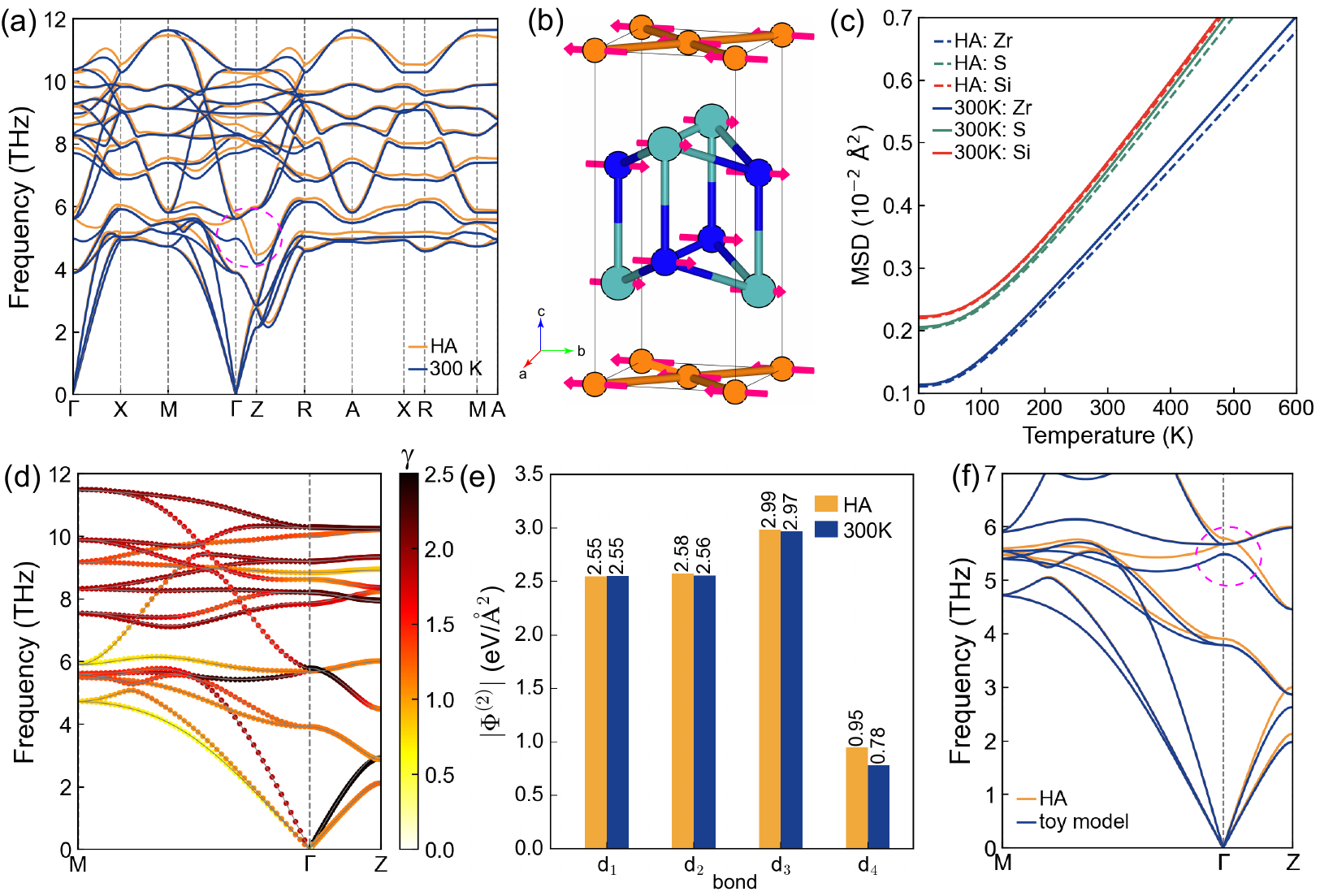}
	\caption{(a) Phonon spectra of ZrSiS obtained from harmonic approximation (HA) and room temperature (300 K) second-order IFCs. The pink dotted line marks the phonon avoided-crossing characteristic. (b) Phonon vibration mode of the sixth phonon branch at $\Gamma$ point ($\Gamma_{6}$) in the first BZ. (c) Mean square displacement (MSD) of Zr, S, and Si obtained using HA and RT second-order IFCs. (d) Phonon dispersion at 0 K with Gr$\rm{\ddot{u}}$neisen parameters projection. (e) Modulus of second-order IFC ($|\rm{\Phi^{(2)}}|$) in ZrSiS. Bonds from d$_{1}$ to d$_{4}$ are defined in Fig.~\ref{fig1}(a). (f) Compare the phonon dispersion relation obtained from our toy model with that calculated using the HA.}
\label{fig2}
\end{figure*}

Figure~\ref{fig2}(a) shows the phonon dispersions along high-symmetry paths in the irreducible Brillouin zone (BZ), calculated with and without inclusion of APR at RT. It can be seen that incorporating anharmonic corrections into the phonon spectrum induces apparent softening of optical phonon modes, while the acoustic phonon dispersions below 4 THz remain almost unchanged. Particularly noteworthy is the emergence of a distinct avoided-crossing behavior between optical modes near 6 THz along the $\bf \Gamma$-{\bf Z} direction (parallel to the $c$-axis, as illustrated in the BZ shown in Fig.~\ref{fig1}(b)). A similar phenomenon was also observed in other systems, such as filled skutterudites \cite{yang2016tuning} and heavy-band NbFeSb compound \cite{han2023strong}, which was found to stem from the rattling motion of guest fillers or heavy element doping. By projecting the phonon density of states (DOS) into different atom sites (see Fig. S4 in SM), we find that these avoided-crossing-related modes are governed by the vibrations of Zr atoms.

Further analysis on the vibrational mode of the sixth phonon branch near the $\bf \Gamma$ point  (denoted as $\Gamma_6$ mode) reveals that APR-induced avoided-crossing behavior is associated with two distinct atomic vibration patterns: (i) cooperative displacements of Zr and S atoms along the $ab$ plane and (ii) counter-directional motion of the Si sublattice, as illustrated in Fig.~\ref{fig2}(b). Due to the relatively weak interlayer atomic bonding, this vibrational feature resembles the behavior of rattling phonon modes in cage compounds, where the relative motion amplitude of atoms between different layers is large, thereby inducing strong anharmonicity. As evidenced by the calculated mean square displacement (MSD) for different atoms in Fig.~\ref{fig2}(c), the TDEP method yields noticeably larger MSD values compared to those obtained from the HA at higher temperatures, particularly for Zr atoms. This indicates that $T$-induced APR effect obviously enhances the lattice anharmonicity associated with Zr atoms, which in turn drives the observed phonon softening around the $\Gamma_6$ mode.   

To clarify why the APR effect exerts the most pronounced softening effect on the phonon modes along the $\bf \Gamma$-{\bf Z} path, we have projected the mode-resolved Grüneisen parameter onto the phonon spectrum in Fig.~\ref{fig2}(d). The Grüneisen parameter serves as an indicator of phonon anharmonicity, with larger values signifying stronger anharmonicity \cite{klemens1951thermal,ziman2001electrons}. As can be seen from the figure, the longitudinal acoustic (LA) branch and the two optical branches that cross near the $\bf \Gamma$ point at about 6 THz display significantly larger Grüneisen parameters than the other branches. This observation aligns with the notable discrepancies in the predicted energies of these specific modes between the HA and TDEP methods, as highlighted in Fig.~\ref{fig5}(a). As the $T$ increases, the anharmonicity of these modes further intensifies, as revealed by the MSD, consequently leading to the softening of these phonon modes and the occurrence of avoided crossing. The increase in anharmonicity is closely related to the weakening of atomic bonding. To explore the evolution of atomic bonding strength with increasing $T$, we compare the modulus of second-order IFCs ($|\rm{\Phi^{(2)}}|$) for the nearest atom pairs with and without APR. As evident from Fig.~\ref{fig2}(e), $|\rm{\Phi^{(2)}}|$ for $d_1$, $d_2$, and $d_3$ remains almost unchanged when temperature effects are taken into account, while a pronounced reduction is observed in $|\rm{\Phi^{(2)}}|$ for $d_4$. This indicates that the Zr-S ($d_4$) bonds weaken with increasing temperature, giving rise to strong anharmonic effects that consequently lead to the observed phonon softening and avoided-crossing behavior.    

To further validate this judgement, we reconstruct the second-order IFCs by artificially reducing the IFC components between Zr-S atomic pairs. The specific implementation includes two steps: (i) reduce the $\Phi_{ij, j\neq i}$ for Zr-S pairs by a scaling factor of 0.5, i.e., $\frac{1}{2}$$\Phi_{ij, j\neq i}$ ($i$ and $j$ belongs to Zr-S atom pairs); (ii) reconstruct the $\Phi_{ii}$ by calculating $-\sum \Phi_{ij, j\neq i}$ to satisfy the acoustic sum rule. The resulting phonon spectrum is shown in Fig.~\ref{fig2}(f). As expected, similar phonon softening and avoided-crossing phenomenon occurs along the $\bf \Gamma$-{\bf Z} path. Hence, the fundamental origin of the avoided-crossing behavior can be attributed to the $T$-induced weakening of Zr-Si atomic bonding. From the perspective of thermal transport, the observed avoided-crossing behavior can affect the phonon group velocity, phonon scattering phase space, and in turn the lattice thermal conductivity \cite{han2023strong,christensen2008avoided,delaire2011giant}.

\subsection{Impact of anharmonic phonon renormalization on lattice thermal conductivity} 
\begin{figure*}[!tbp]
	\includegraphics[width=\linewidth]{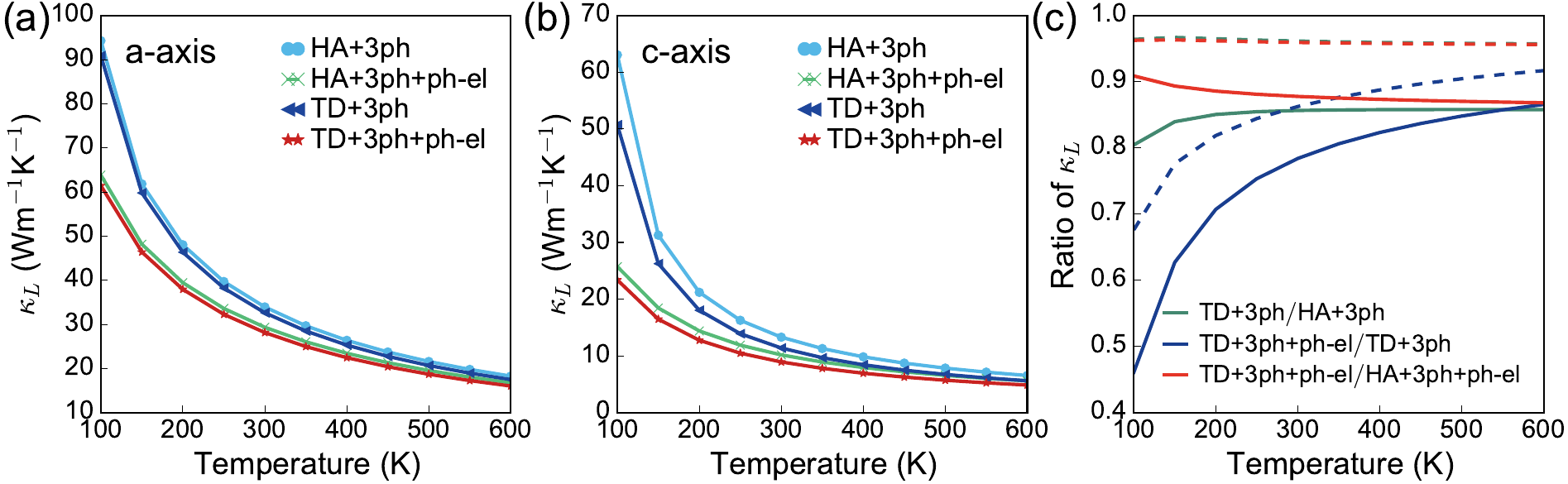}
	\caption{The lattice thermal conductivity ($\kappa_{\rm L}$) of ZrSiS along the (a) $a$-axis and (b) $c$-axis, as well as (c) the ratio of $\kappa_{\rm L}$, calculated using different computational approaches. The dotted lines in (c) denote the corresponding ratios of $\kappa_{\rm L}$ along the $a$-axis.}
\label{fig3}
\end{figure*}

Having established a basic understanding of APR effects on the lattice dynamics in ZrSiS, we then investigate its lattice thermal conductivity using different levels of theory. As seen in Fig.~\ref{fig3}(a, b), the APR effect has a very weak influence on the $\kappa_{\rm L}$ along the $a$-axis, but significantly reduces the $\kappa_{\rm L}$ along the $c$-axis over the entire $T$ range. This observation is consistent with the fact that the APR effect manifests its most pronounced impact on phonon modes along the $\bf \Gamma$-{\bf Z} path (corresponding to the $c$-axis). Moreover, our calculation shows that ph-el scattering plays a crucial role in reducing the $\kappa_{\rm L}$ along both $a$ and $c$ axes, particularly at lower temperatures. With considerations of renormalization effects and ph-el scattering, the predicted $\kappa$ along the $a$-axis is 28 W/mK at RT, which is more than three times higher than that along the $c$-axis (9 W/mK). This strong anisotropy in $\kappa_{\rm L}$ stems from the bonding heterogeneity between interlayer and intralayer atomic pairs as discussed above. Note that the effect of higher-order 4ph scattering on $\kappa_{\rm L}$ is found to be negligible, as demonstrated in Fig. S5 of the SM \cite{SM}, and is thus not considered in the subsequent calculation.

To highlight roles of APR effect and ph-el scattering, we present the ratio of $\kappa_{\rm L}$ between calculations with and without incorporating these effects in Fig.~\ref{fig3}(c). By comparing the $\kappa_{\rm L}$ values obtained by the HA+3ph model and the TD+3ph model, we find that APR leads to a marginal reduction of less than 4$\%$ along the $a$-axis, but induces a noticeable suppression of up to 16$\%$ along the $c$-axis at temperatures above RT. By contrasting the predictions from the TD+3ph+ph-el model and the TD+3ph model, we reveal that ph-el scattering causes the $\kappa_{\rm L}$ along the $a$-axis to decrease by 15$\%$ at RT and by more than 30$\%$ at 100 K. This strong effect arises because the ph-el scattering rate is nearly $T$-independent \cite{yang2021tuning}, whereas the 3ph scattering rate scales approximately linearly with $T$. As a result, the suppression of $\kappa_{\rm L}$ by ph-el scattering is most pronounced at low and intermediate temperatures. Most strikingly, the suppression is even stronger along the $c$-axis, with reductions of 25\% at RT and over 50\% at 100 K. Further comparing the results obtained from the TD+3ph+ph-el model and the HA+3ph+ph-el model, we can see that the APR effect can reduce the $\kappa_{\rm L}$ along the $c$-axis by more than 12\% even at RT. These findings explicitly reveal the critical importance of APR effect and ph-el interactions in determining the $\kappa_{\rm L}$ of ZrSiS, especially along the $c$-axis.   

\begin{figure}[!tbp]
	\includegraphics[width=\linewidth]{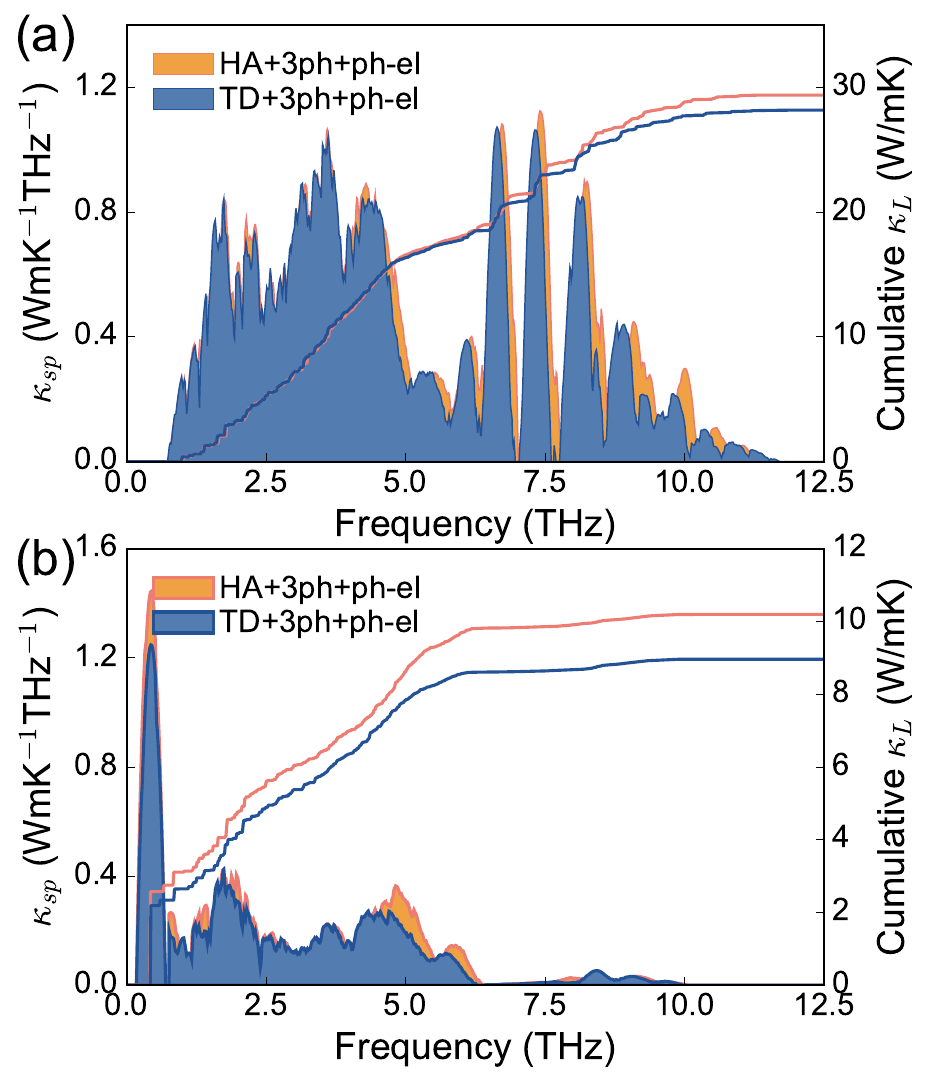}
	\caption{The spectral $\kappa_{\rm L}$ ($\kappa_{\rm sp}$) and cumulative $\kappa_{\rm L}$ of ZrSiS along (a) $a$-axis and (b) $c$-axis using HA and TD IFCs, respectively.}
\label{fig4}
\end{figure}

For a clearer insight into the suppression of $\kappa_{\rm L}$ by the APR, we compare the spectral $\kappa_{\rm L}$ and its accumulation with phonon frequency obtained using the HA+3ph+ph-el model and the TD+3ph+ph-el model in Fig.~\ref{fig4}(a). It is found that along the $a$-axis, optical phonons make a significant contribution ($\sim$40\%), whereas along the $c$-axis, acoustic phonons dominate thermal transport ($>$95\%). Furthermore, the predicted cumulative $\kappa_{\rm L}$ from the two models along the $a$-axis begins to differ slightly only at frequencies above 6 THz, which is mainly due to the softening of high-frequency optical phonons by the APR effect. In contrast, it can be seen from Fig.~\ref{fig4}(b) that a relatively significant difference is observed along the $c$-axis within the frequency ranges of 0–3 THz and 4–6 THz. This implies that the APR-induced phonon avoided crossing could considerably suppress the $\kappa_{\rm L}$ along the $c$-axis. 

\begin{figure*}[!tbp]
	\includegraphics[width=\linewidth]{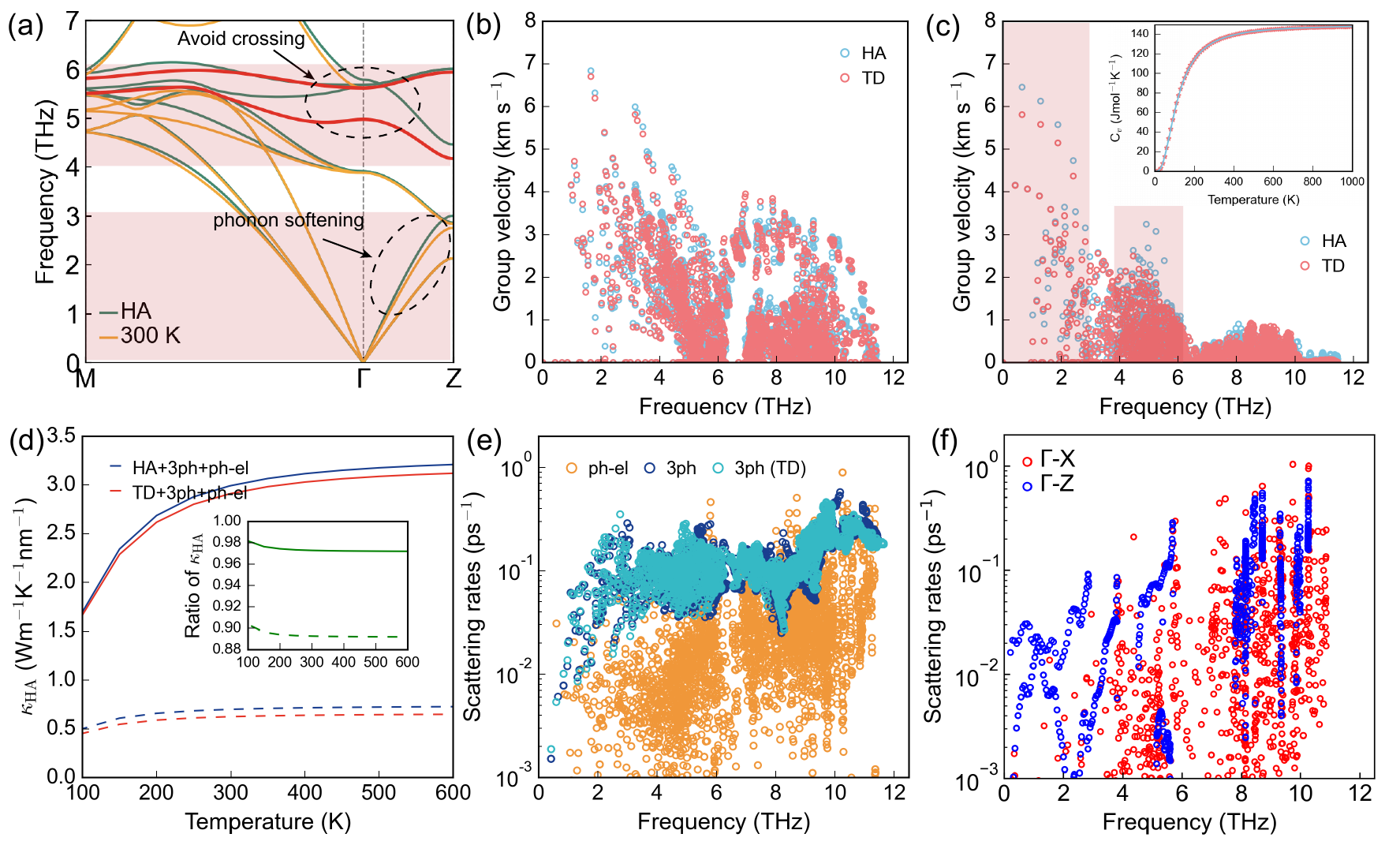}
	\caption{(a) The zoom-in plot of phonon dispersions highlighting the phonon softening and avoided crossing behavior. Phonon group velocity of ZrSiS along the (b) $a$-axis, and (c) $c$-axis. (d) The harmonic component of $\kappa_{\rm L}$ ($\kappa_{\mathrm{HA}}$), with the inset illustrating the ratio of $\kappa_{\mathrm{HA}}$ obtained with and without using TD IFCs. (e) Three-phonon (3ph) scattering rates obtained with and without using HA and TD IFCs, as well as the ph-el scattering rates at 300 K. (f) The path-dependent ph-el scattering rates.}
\label{fig5}
\end{figure*}

To unravel the microscopic mechanism behind the suppressed $\kappa_{\rm L}$, we further analyze phonon specific heat ($C_{p \mathbf{q}}$), phonon group velocity ($v_{p \mathbf{q}}$), and phonon lifetime ($\tau_{p \mathbf{q}}$). As emphasized in Fig.~\ref{fig5}(a), the renormalized phonon spectrum exhibits significant band avoided-crossing behavior within 4-6 THz, along with a noticeable softening of the LA branch below 3 THz. These frequency ranges align closely with the regions where the distinct variations in $\kappa_{\rm L}$ along the $c$-axis are observed. The direct consequence of these alterations in the phonon spectrum is the deceleration of heat-carrying phonons. As illustrated in Figs.~\ref{fig5}(b) and (c), the incorporation of the APR effect leads to only minor variations in the modal group velocity along the $a$-axis. In contrast, the modal group velocity along the $c$-axis undergoes a considerable reduction within the frequency ranges of 0-3 THz and 4-6 THz. Besides, the inset of Fig.~\ref{fig5}(c) reveals that the specific heat derived from the HA and TD methods show complete coincidence over the entire $T$ range. To quantify the impact of APR on specific heat and phonon group velocity, we calculate the harmonic component of thermal conductivity ($\kappa_{\rm HA}$), defined as $\kappa_{\rm HA}=\sum_{p \mathbf{q}}C_{p \mathbf{q}}v_{p \mathbf{q}}^{\alpha}v_{p \mathbf{q}}^{\beta}$. Fig.~\ref{fig5}(d) displays a comparison of $\kappa_{\rm HA}$ calculated using the HA and TD methods for $a$ and $c$ axes, respectively, which suggests that the APR effect more or less reduces the harmonic component of $\kappa_{\rm L}$ along both axes. From the relative changes in $\kappa_{\rm HA}$ shown in the inset of Fig.~\ref{fig5}(d), we see that the reduction along the $a$-axis is less than 3\% above RT, whereas the reduction along the $c$-axis is as much as 12\%. These changes are quantitatively consistent with the $\kappa_{\rm L}$ ratios depicted in Fig.~\ref{fig3}(c). Lastly, we examine the influence of APR on the 3ph scattering rates. It is seen from Fig.~\ref{fig5}(e) that the inclusion of APR results in only minor changes in the 3ph scattering rates throughout the frequency range. Combining these results, we conclude that the APR-induced reduction in $\kappa_{\rm L}$ along the $c$-axis is attributed to the depressed group velocities of heat-carrying phonons, resulting from phonon softening and avoided crossing behavior.

Also shown in Fig.~\ref{fig5}(e) is the comparison between ph-el and 3ph scattering rates at RT, which indicates that the former is substantially weaker than the latter, particularly for the lower-frequency phonons that carry most heat. Despite its relatively weak magnitude, the ph-el scattering still induces a significant reduction in $\kappa_{\rm L}$, with an especially pronounced effect along the $c$-axis, as revealed above. To elucidate the more pronounced suppression effect of ph-el scattering on $\kappa_{\rm L}$ along the $c$-axis compared to the $a$-axis, we depict the path-dependent ph-el scattering rates in Fig.~\ref{fig5}(f). Notably, for phonons with frequencies below 6 THz, which contribute the most to the $c$-axis $\kappa_{\rm L}$ as illustrated in Fig.~\ref{fig4}(b), the ph-el scattering rates along the $\bf \Gamma$-{\bf Z} path are considerably higher than those along the $\bf \Gamma$-{\bf X} path, the latter corresponding to the $a$-axis. This disparity underscores a stronger ph-el interaction along the $c$-axis relative to the $a$-axis, directly accounting for the more significant reduction in $\kappa_{\rm L}$ observed along the $c$-axis. The strength of ph-el scattering is known to be positively correlated with the electronic DOS at the Fermi level \cite{kundu2021ultrahigh}. In ZrSiS, the Zr 4$d$ orbitals contribute significantly to the electronic DOS at the Fermi level (see Fig.~\ref{fig6}(a)), thereby leading to relatively strong ph-el effect compared to other semimetals like ScZn \cite{ding2024anharmonicity}.

\subsection{Anomalously large lattice contribution to thermal conductivity and Lorenz number} 
\begin{figure*}[!t]
	\includegraphics[width=\linewidth]{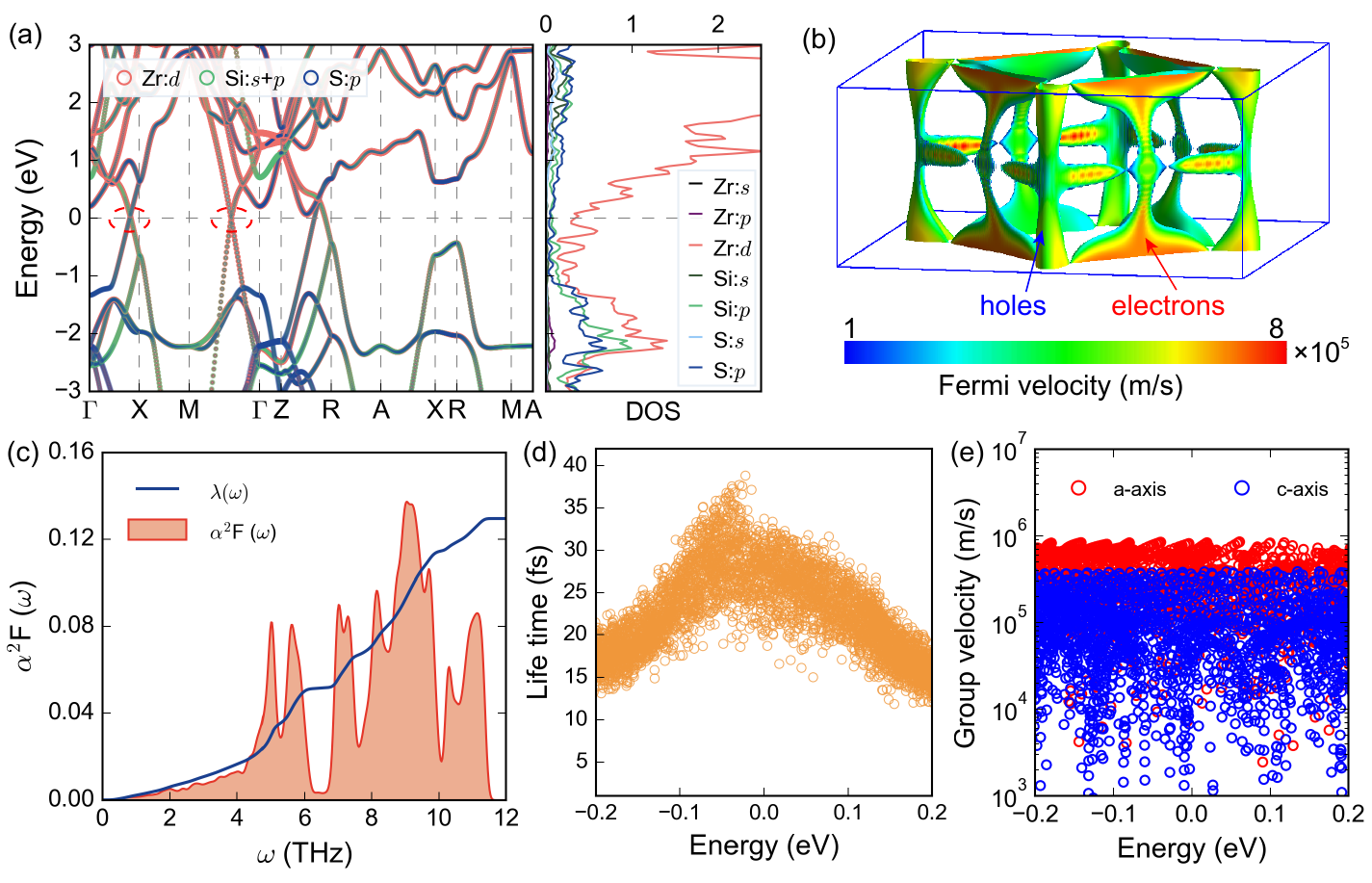}
	\caption{(a) Projected electronic band structure and DOS of ZrSiS. The dotted red circles denote the Dirac points. (b) Fermi surface with fermi velocity projection. (c)  Eliashberg spectral function $\alpha^{2}$F($\omega$) and electron-phonon coupling strength $\lambda$($\omega$). (d) Phonon-limited carrier lifetimes. (e) Carrier group velocities along the $a$ and $c$ axes.}
\label{fig6}
\end{figure*}

We then turn attention to the electronic contribution to $\kappa$, which is closely linked to the electronic band structure. Figure~\ref{fig6}(a) shows the orbital decomposed electronic band structure together with the projected DOS. It is clear that the electronic states near the Fermi level are dominated by the Zr 4$d$ orbitals. Notably, the distinct Dirac cones are observed along the $\bf \Gamma-X$ and $\bf \Gamma-M$ paths as marked by circles, which contribute to both high Fermi velocities and weak el-ph coupling and thus may yield superior charge transport. These Dirac cones are protected by the $C_{2v}$ symmetry, and the spin orbital coupling effect was found to be minimal and only produce minor gaps (roughly 20 meV) \cite{schoop2016dirac}. Therefore, the massless linear dispersions are preserved. 

The Fermi velocities ($v_{\rm F}$) of ZrSiS are mapped onto its Fermi surface in Fig.~\ref{fig6}(b), revealing exceptionally high electron and hole velocities up to 8$\times10^5$ m/s, which is a direct consequence of the linear Dirac band dispersions. Also, ZrSiS displays significant differences in Fermi velocities along different crystallographic axes owing to the lattice anisotropy, as evidenced in Fig.~\ref{fig6}(e). On the other hand, the Eliashberg spectral function $\alpha^2F(\omega)$ presented in Fig.~\ref{fig6}(c) reveals weak el-ph coupling of ZrSiS across the entire phonon frequency range, with a coupling strength $\lambda$ of merely 0.13, which is substantially lower than that of most known metallic compounds \cite{tong2019comprehensive}. The weak el-ph coupling can be connected to the featured Dirac cones, since they account for small Fermi pockets and consequently a markedly low electronic DOS near the Fermi level. Using the computed $\lambda$ value, we estimate the carrier lifetime of ZrSiS through the relation $\tau=\hbar(2\pi k_{\rm B}T\lambda)^{-1}$ \cite{allen1978new,ding2025concurrent}, obtaining a value of $\sim$31 fs that yields a close agreement with our DFT calculation, as demonstrated in Fig.~\ref{fig6}(d). 

\begin{figure*}[!t]
	\includegraphics[width=\linewidth]{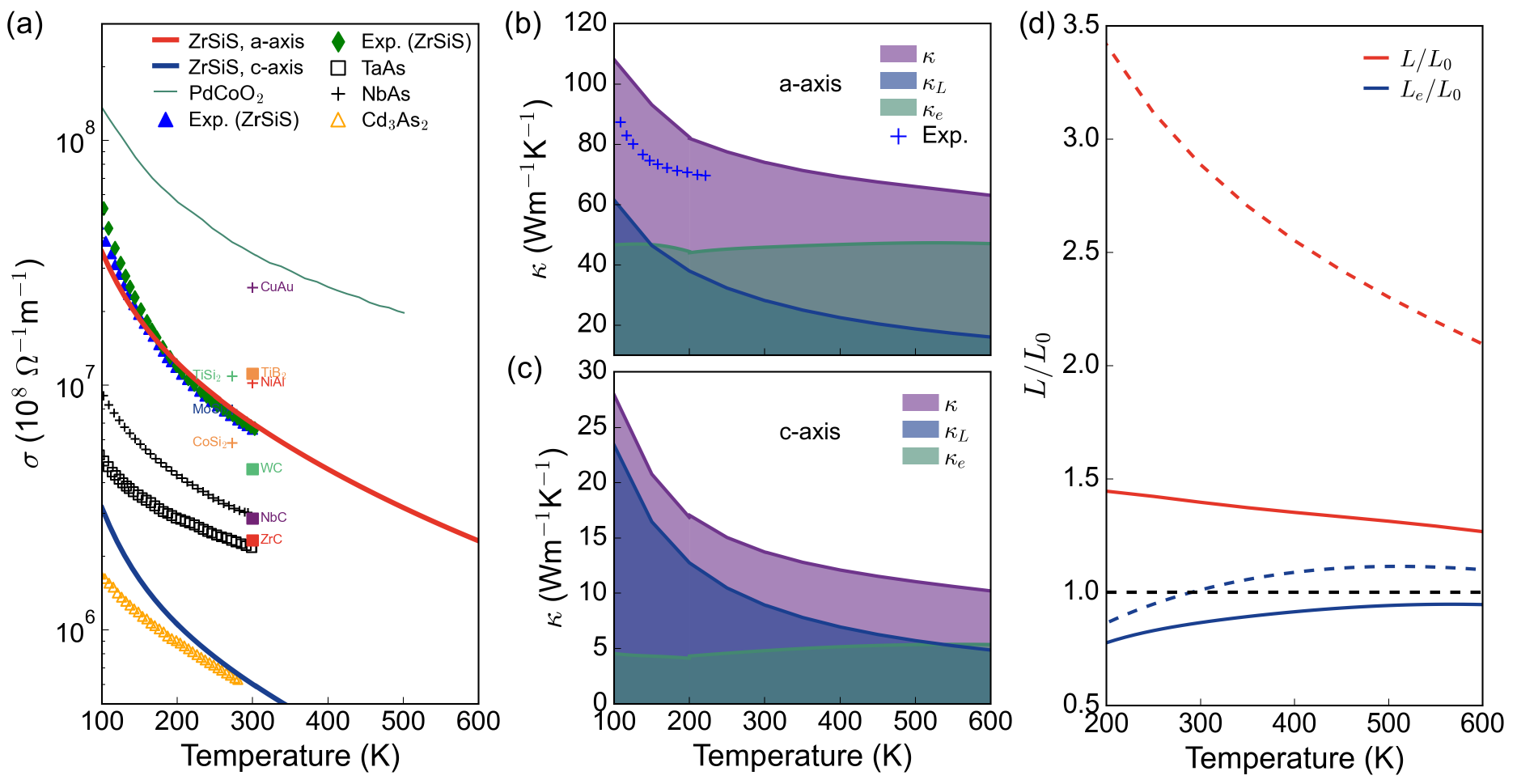}
	\caption{(a) Electrical conductivity of ZrSiS. The experimental data for ZrSiS are represented by blue triangles \cite{hussain2020electron} and green diamonds \cite{singha2017large}. The experimental data for TaAs \cite{xiang2017anisotropic}, NbAs \cite{zhang2019ultrahigh}, Cd$3$As$2$ \cite{he2014quantum}, XC (X = Zr, W, Nb) \cite{he2014quantum}, and CoSi$2$ \cite{hirano1990electrical} are also included for comparison. (b, c) Lattice thermal conductivity ($\kappa_{\rm L}$), electronic thermal conductivity ($\kappa_{\rm e}$), and total thermal conductivity ($\kappa$) along the a and c-axes, respectively. Green plus symbols represent the experimental data \cite{hussain2020electron} for the total $\kappa$ along the a-axis of ZrSiS. (d) Lorenz number as a function of temperature.}
\label{fig7}
\end{figure*}

The high Fermi velocity and long carrier lifetime are conducive to superior electrical conductivity. The calculated phonon-limited electrical conductivity $\sigma$ as a function of $T$ along different axes is plotted in Fig.~\ref{fig7}(a), and values along the $a$-axis agree well with the experimental data \cite{hussain2020electron,singha2017large}. As expected, ZrSiS exhibits excellent conductivity, with the $\sigma$ along the $a$-axis reaching 6.7$\times$$10^{6}$ $\Omega^{-1}\rm{m}^{-1}$ at RT. This value, although lower than that of well-known high-$\sigma$ intermetallic compounds like PdCoO$_{2}$ \cite{williams1999electrical}, significantly surpasses those of many other topological semimetals and related compounds such as TaAs \cite{xiang2017anisotropic}, NbAs \cite{zhang2019ultrahigh}, Cd$_{3}$As$_{2}$\cite{he2014quantum}, XC (X = Zr, W, Nb) \cite{he2014quantum}, and CoSi$_{2}$ \cite{hirano1990electrical}. In addition, we can see that the $\sigma$ of in ZrSiS along the $a$-axis is nearly one order of magnitude higher than that along the $c$-axis, which is attributed to the distinct Fermi velocities between two directions as discussed above.

We then predict the electronic thermal conductivity ($\kappa_{\rm e}$) of ZrSiS using Eq.\ref{kappae} and compare it with the lattice contribution in Figs.~\ref{fig7}(b) and (c). It is seen that the calculated $\kappa_{\rm e}$ shows the behavior of typical metal with a weak $T$ dependence. Along the $a$-axis, $\kappa_{\rm e}$ contributes the majority of thermal conductivity above 150 K, despite the non-negligible contribution from the lattice counterpart. Taking both contributions into account, the predicted total thermal conductivity, $\kappa=\kappa_{\rm e}+\kappa_{\rm L}$, is displayed in Fig.~\ref{fig7}(b), with the intrinsic upper limit reaching 74 W/mK at RT. It should be noted that the calculated $\kappa$ is considerably lower than the measured data within 100–200 K \cite{hussain2020electron}. This discrepancy can be reasonably attributed to the presence of defects or impurities in experimental samples \cite{hussain2020electron,singha2017large}, which introduce additional scattering centers for heat-carrying phonons and thus noticeably suppresses thermal transport, especially at lower temperatures. Along the $c$-axis, we find, surprisingly, that the lattice contribution to $\kappa$ is substantially largely than the electronic contribution below 500 K. This behavior is anomalous since the $\kappa_{\rm e}$ in metallic systems typically dominates the heat transport. 

In experiments, the Lorenz number $L$ serves as a crucial parameter connecting thermal conductivity and electrical conductivity ($\sigma$) via the WFL. However, as documented in the literature \cite{berman1978thermal}, the actual $L$ in metals may deviate from the Sommerfeld value $L_{0}$ and exhibit strong temperature dependence. Figure~\ref{fig6}(d) shows the calculated $T$-dependent $L$ of ZrSiS scaled by $L_0$. Considering only the electronic contribution to $\kappa$, the predicted Lorenz number, $L_e=\kappa_{\rm e}/\sigma T$, increases with temperature and goes toward $L_0$. However, when adding the lattice contribution, the predicted $L$ values along both axes show substantial deviations from $L_{0}$ and exhibit a pronounced $T$ dependence. More specifically, the anomalously large $\kappa_{\rm L}$ leads to an approximately 1.5-fold increase in the $a$-axis $L$ relative to $L_0$ near RT, while the $L$ along the $c$-axis even exhibits a nearly 3-fold enhancement. These results explicitly suggest that the conventional Sommerfeld value inadequately describes the coupled thermal and electrical transport behavior in ZrSiS, especially along the $c$-axis. %highlighting the importance of separating electronic and lattice contributions to $\kappa$ in topological semimetals.

\section{Summary}
In summary, our first-principles calculations establish a unified framework for understanding both electrical and thermal transport in ZrSiS. The APR effect is found to cause noticeable phonon softening, leading to substantial suppression of $\kappa_{\rm L}$. Most notably, the $T$-induced weakening of Zr-S atomic interactions induces the avoided crossing in low-frequency optical phonon modes. The synergistic combination of this avoided-crossing behavior and phonon softening effectively decreases the group velocity of heat-carrying phonons, ultimately resulting in a 16\% reduction of $\kappa_{\rm L}$ along the $c$-axis. Moreover, we find that the lattice contribution to $\kappa$ is anomalously large compared to its electronic counterpart, and it persists as the dominant component along the $c$-axis despite the considerable suppression of $\kappa_{\rm L}$ by ph-el scattering at low and intermediate temperatures. Consequently, the resulting Lorenz number deviates significantly from the expected Sommerfeld value by up to a factor of three. Additionally, our calculation reveals that ZrSiS exhibits exceptional electrical conductivity due to high Fermi velocities and weak el-ph coupling, originating from its topological Dirac states near the Fermi level. These findings provide fundamental insights into coupled electrical and phonon transport in ZrSiS and are expected to stimulate further exploration of thermal transport in topological semimetals.

\section{Acknowledgement}
This work was supported by the National Natural Science Foundation of China (Grants No.~12404045, No.~12374038, No.~12147102, and No.~52371148), the Science and Technology Research Program of Chongqing Municipal Education Commission (Grants No. KJZD-K202500512 and No. KJQN-202400553) and the Natural Science Foundation of Chongqing (Grants No. CSTB2025NSCQ-GPX1028, No. CSTB2023NSCQ-LZX0138, No. CSTB2023NSCQ-JQX0024, and No. CSTB2022NSCQ-MSX0834).

% \bibliography{ref/reference}


~~~\\
~~~\\
~~~\\

\end{document}